\documentclass[10pt,conference]{IEEEtran}
\IEEEoverridecommandlockouts
\usepackage{cite}
\usepackage{amsmath,amssymb,amsfonts}
\usepackage{algorithmic}
\usepackage{graphicx}
\usepackage{textcomp}
\usepackage{xcolor}
\usepackage{url}
\usepackage{multirow}
\usepackage{makecell}
\usepackage{graphics}
\usepackage{enumitem}
\usepackage{etoolbox}
\usepackage{color, colortbl}
\usepackage{lipsum}
\usepackage{tabularx,booktabs}
\usepackage[section]{placeins}
\usepackage{graphicx}
\usepackage{comment}
\usepackage{floatrow}
\usepackage[para]{footmisc}
\usepackage{rotating}
\usepackage{tablefootnote}
\restylefloat{table}
\usepackage{breakurl}
\usepackage[nobiblatex]{xurl}
\definecolor{lightgray}{rgb}{0.83, 0.83, 0.83}
\def\BibTeX{{\rm B\kern-.05em{\sc i\kern-.025em b}\kern-.08em
    T\kern-.1667em\lower.7ex\hbox{E}\kern-.125emX}}
\begin{document}

\title{Saudi Arabian Perspective of Security, Privacy, and Attitude of Using Facial Recognition Technology}

\author{\IEEEauthorblockN{Amani Mohammed Alqarni}
\IEEEauthorblockA{\textit{Dept. of Computer Science} \\
\textit{California State University San Marcos}\\
San Marcos, United States \\
alqar003@csusm.edu}
\and
\IEEEauthorblockN{Daniel Timko}
\IEEEauthorblockA{\textit{Dept. of Computer Science} \\
\textit{California State University San Marcos}\\
San Marcos, United States \\
timko002@csusm.edu}
\and
\IEEEauthorblockN{Muhammad Lutfor Rahman}
\IEEEauthorblockA{\textit{Dept. of Computer Science} \\
\textit{California State University San Marcos}\\
San Marcos, United States \\
mlrahman@csusm.edu}}

\maketitle

\begin{abstract}
Facial Recognition Technology (FRT) is a pioneering field of mass surveillance that sparks privacy concerns and is considered a growing threat in the modern world. FRT has been widely adopted in the Kingdom of Saudi Arabia to improve public services and surveillance. Accordingly, the following study aims to understand the privacy and security concerns, trust, and acceptance of FRT in Saudi Arabia. Validated Privacy Concerns (IUIPC-8), Security Attitudes (SA-6), and Security Behavior (SeBIS) scales are used along with replicate studies from Pew Research Center trust questions and government trust questions. In addition, we examine potential differences between Saudis and Americans. To gain insights into these concerns, we conducted an online survey involving 53 Saudi Arabia citizens who are residing in the USA. We have collected data in the US instead of Saudi Arabia to avoid the regulatory challenges of the Saudi Data \& Artificial
Intelligence Authority (SDAIA). Responses from closed-ended questions revealed that Saudis score much lower than Americans when it comes to security attitudes, whereas they score lower when it comes to privacy concerns. We found no significant difference between Saudis' and Americans' acceptance of the use of FRT in different scenarios, but we found that Saudis trust advertisers more than Americans. Additionally, Saudis are more likely than Americans to agree that the government should strictly limit the use of FRT.
\end{abstract}


\section{Introduction}
Facial Recognition Technology is rapidly becoming widespread in the global context. 
It is estimated that by 2021 over 1 billion security cameras have been installed for both private and public purposes~\cite{comparitech}.
China has the largest video surveillance network in the world and has now nearly one billion surveillance cameras~\cite{chinafrtadopt}. The East Asia/Pacific and the Middle East/North Africa regions are active adopters of facial recognition and other identity tools. South and Central Asia and the Americas also demonstrate sizable adoption of AI surveillance instruments. As FRT usage has grown rapidly, personal information disclosure and data collection has increased; however, providing societal safety while balancing individual privacy rights has been challenging ~\cite{lillo2014open}. FRT raises more anxiety than other identity methods, such as fingerprint and iris recognition, because it can be employed anytime without users noticing. Therefore, it is unsurprising that there are many concerns associated with it. It is slowly making an appearance within different applications like education, retail, access control to certain internet of things (IoT) devices, transportation, hospitality, and banking.

As of now, private and public FRT are creating an unprecedented dilemma. If we connect private and public FRT to a network, where data can be continuously obtained easily and combined with databases of public information, that could enable automated identification and tracking of people. A very early version of FRT was tested at the 2002 Super Bowl when law enforcement officials scanned people in the crowd without their permission and found several minor criminals. However, the experiment was subject to a high number of false positives. Following that test, FRT blossomed in the 2010s due to rapid developments in artificial intelligence ~\cite{batra2018artificial}. The increasing use of facial recognition systems has yielded ample research opportunities as well as many potential uses and benefits in the public and private sector. Nonetheless, there are still significant risk perceptions that accompany the adoption of this technology~\cite{zhang2021facial,pubattitudescriminaljustice}.
To address the substantial security and privacy concerns, it is important to know whether there are meaningful differences in privacy and security preferences, beliefs, and attitudes between people of different nations. For example, a study has been done to compare the security and convenience between four nations~\cite{kostka2021between}. Attitude studies on specific applications that use FRT have also been done ~\cite{liu2021resistance,yang2021attitudes,ritchie2021public}. Recent studies have been done on adults in the United States to discover their trust level toward law enforcement’s use of FRT~\cite{usfrt}. Results revealed that more than 50\% of U.S. participants trust law enforcement to use FRT. In contrast, when used by advertisers or technology companies, Americans are less accepting of facial recognition software. Most of the participants in the study by Smith et al.~\cite{smith2019more} believe that FRT can effectively and easily identify individual people and classify them by gender and race. 

With Vision 2030, Saudi Arabia is opening to the world with a unique blueprint for economic and social reform ~\cite{grand2020assessing}. This includes one of the most vital components of its transformation program: digital transformation. By expediting the implementation of primary and digital infrastructure projects, the program aims to increase operational excellence in government, improve economic enablers, and enhance living standards. Artificial intelligence-based technology has been deployed more widely due to digital transformation. In the Fall of 2022, the Kingdom of Saudi Arabia approved the use of security surveillance cameras in the country ~\cite{saudicabinet}. As a result, approximately 22 places, including schools, universities, hospitals, clinics, medical cities, and private health facilities, were required to install surveillance cameras. 

To address the security and privacy concern, there are numerous studies have been conducted on the use of FRT. However, most research conducted on privacy concerns regarding FRT is focused on Western cultures~\cite{steinacker2020facial, katsanis2021survey, zhang2021facial, ritchie2021public, kostka2022under, smith2019more} and very little is conducted in the Middle East. The behaviors people engage in regarding their privacy are firmly rooted in various cultural beliefs and values ~\cite{li2022cultural}. In Muslim countries, women dress differently in public \cite{poushter2014people}. As a Muslim majority country, Saudi Arabia has unique culture and norms. The majority of women cover their faces in Saudi Arabia. Consequently, it is more likely that men's identity information will be collected publicly than women's in a place like Saudi Arabia. Given the unique context of Saudi Arabians, our study focuses on the perspectives of Saudi Arabian citizens residing in the united states regarding the use FRT for surveillance purposes.

\textbf {In the following paper, we explore privacy and security concerns for Saudi Arabia by addressing the following research questions:}

\textbf {RQ1:} {\emph Are there meaningful differences in security, privacy concern, and security behaviors?}

\textbf {RQ2:} {\emph Are there meaningful differences in public acceptance when using FRT?}

\textbf {RQ3:} {\emph Are there meaningful differences in opinions between locals of Saudi and the US in regard to FRT? And to what extent do gender and age impact security behaviors and privacy concerns for Saudis?}

To address these questions, we have administered an online survey to 53 Saudi Arabian citizens residing in the United States. These participants were recruited through word-of-mouth referrals and from online messaging channels for English-speaking Saudi groups in America. We recorded participants' responses to a seven-part questionnaire including questions concerning participant demographics, behavioral attitudes and privacy concerns, and public awareness and potential misuse of FRT. Additionally, we reused questions from a Pew Research Center and Center for Data Innovation survey to compare our responses to the general American public. Our research shows differences in security and privacy concerns related to FRT use in several surveillance scenarios. 

\textbf{Our Contributions:} To understand Saudis' complaints, we analyze their concerns, opinions, awareness, and beliefs, and how they are different between the US populations. \textbf{Our work made the following contributions:}

\begin{enumerate}[leftmargin=*]
   \item To our best knowledge, this is the first quantitative study that measures the Saudi Arabian perspective of security and privacy attitudes and concerns regarding FRT.  
   \item We compared the security perspectives of Saudi Arabians with Americans to determine similarities and differences between both groups. 
 \end{enumerate}

\textbf{Summary of Key Results:} 
This research provided following key insights:

\begin{enumerate}[leftmargin=*]
   \item We confirm that the accuracy of FRT  has an impact on the  Saudi Arabian populations' support for the technology. When FRT use is proposed for personal identification with varying degrees of accuracy, we saw that both Saudi and US approval consistently increased with the higher accuracy. The difference between 80\% accuracy and 100\% accuracy in identifying suspects using FRT resulted a 17\% increase in approval of the use of FRT technology by police. 
   \item Our results show that Saudi Arabian participants have a higher propensity towards government limitation of surveillance. Across all propositions, we saw a consistently higher rate of agreement in the proposition that FRT should be limited by the police and government.
   \item Our analysis of security behaviors shows that the Saudi population has a low average acceptance level of FRT in our tested scenarios. Additionally, we found that Saudi participants did not believe FRT is accurate in effectively identifying someone's race.
 \end{enumerate}

In the following section, we present work related to FRT vulnerabilities, such as biometric data challenges. We then describe the methods and results of our survey. Finally, we discuss our findings and synthesize conclusions.

\section{Related Work}
FRT has been a focal point of research over the last few decades. In this section, we discuss works related to three key areas: Facial Recognition Technology Vulnerabilities, Biometric Recognition Related Privacy, and Security Concerns, and Public Acceptance of FRT. The implications of FRT increase with increasing self-disclosure on the internet ~\cite{acquisti2014face}. Despite the advantages of FRT surpassing the disadvantages, privacy challenges are the most significant implication of this technology, and many searches have been done to reduce privacy concerns and increase awareness  ~\cite{hamann2019facial, naker2017now}.

\textbf{FRT Vulnerabilities \& Biometric Data Challenges}
Facial recognition is explained as a computer that takes an image and calculates the distance between major structural pieces like the nose and eyes. The facial recognition system uses a template to analyze images of people’s identities ~\cite{Gross2021ValidityAR}. Once a computer recognizes a face, it searches its existing template of images to see if it can locate a matching code. Due to the COVID-19 pandemic, new challenges have emerged that complicate the facial recognition system.
Face masks obstruct a significant part of the face, leading to low recognition performance. This case is not only for impermeable masks but also for transparent face covers because reflections produce a variation that is non-trivial to the model. However, these difficulties can be overcome by focusing on parts of the face that remain uncovered, such as the iris and the wider periocular region of the face ~\cite{gomez2021biometrics}. Many countries have directly collaborated with private companies to develop digital solutions based on their requirements, without the supervision of legislative institutions and in the absence of public discussion. This revealed how these systems fail to meet even the most basic thresholds of legality, proportionality, accountability, necessity, legitimacy, or safeguarding.

Previous studies have discussed the challenges of different biometric technologies. A significant challenge in using biometric recognition systems is to build a robust and appropriate sensor that minimizes recognition error ~\cite{jain201650}. A comparison between different biometric methods has shown that facial recognition biometrics has low accuracy, high cost, large template size, low long-term stability, and low security level~\cite{saini2014comparison}.
One study that presented an analysis of the biometric authentication that causes new challenges to security and privacy discussed the danger of frequent biometric authentication that companies use to identify persons to evaluate their buying decisions. Access to people's identity information will lead to illegal spying from different agencies. As a result, security and privacy concerns will increase over time. However, signal processing is important in providing solutions to decrease security and privacy concerns ~\cite{memon2017biometric}. Another challenge is facial expression bias, which impacts facial recognition systems since nearly all popular FRT databases show massive facial recognition biases ~\cite{pena2021facial}. The authors addressed the FRT challenges that can be concluded in pose variations like variation in lightning conditions and illumination problems which are observed in people who wear collusions like hats and eyeglasses ~\cite{hassaballah2015face,olszewska2016automated,singh2018techniques,6724278}.

\textbf{Biometric Recognition: Privacy and Security Concerns} Since biometric features are not secretive, it is possible to obtain a person’s face biometric without their knowledge. This permits covert recognition of previously recorded people. Consequently, those who yearn to remain anonymous could be denied ~\cite{prabhakar2003biometric}. In this study, a machine learning methodology was presented to efficiently recognize the masked faces, inspired by the state-of-the-art algorithms; the proposed method achieved 99\% accuracy on the classifier, which was built on the masked faces dataset. Due to COVID-19, masked faces have created a considerable challenge for facial recognition. However, this simple yet innovative approach effectively solves the problem and addresses security and social concerns ~\cite{mundial2020towards}. However, there are growing concerns that when COVID-19 ends, data gleaned from these digital systems could be misused. The lack of adequate regulations does not guarantee that governments will restrict their measures, particularly where there is no specific legislation establishing rules concerning the processing, storing, or discarding of the collected data.

\textbf{Public acceptance and discrimination Toward FRT} Using biometric systems for remote detection has raised social, cultural, and legal concerns~\cite{national2010biometric}. In healthcare, FRT poses new challenges regarding privacy~\cite{martinez2019important}. In schools, FRT alters the nature of schools and schooling along divisive, authoritarian, and oppressive lines ~\cite{andrejevic2020facial}. One study compared the privacy concerns within the United States justice system and FRT outside of the United States justice system. Additionally, the author presented the ethical and legal concerns associated with FRT ~\cite{naker2017now}. Similarly, another study discussed privacy concerns across multiple deployment scenarios of facial recognition and strategies for the deployment of facial recognition. Specifically, the author focused on ensuring that people have the same level of transparency and control ~\cite{zhang2021facial}. A high level of general awareness about FRT has been presented. In automobile security, smartphone usage rates are higher in China than in the US ~\cite{kostka2021between}. Also, this study analyzes the interplay between technical and social issues involved in the widespread application of video surveillance for person identification ~\cite{bowyer2004face}. Public attitudes, trust, and familiarity with FRT scenarios have been analyzed ~\cite{lai2021has}. Insight regarding the public’s attitudes from China, the UK, Germany, and the US have been gathered, and the results provided input for policymakers and legal regulations ~\cite{steinacker2020facial}. In several studies, gender has revealed distinctions. More specifically, the author explores whether men and women think differently about privacy, and he found that men are more likely to approve of the use of cameras using FRT in the workplace ~\cite{stark2020don}. If an individual identifies as a white man, then the software is right 99\% of the time, while the accuracy is 35\% when women with darker skin ~\cite{lohr2018facial}. In the workplace, women are more concerned and less likely than men to accept using a camera that has FRT ~\cite{stark2020don}. Gender bias is one of the consequences of using FRT, and in this study, the authors evaluate potential gender bias ~\cite{atay2021evaluation,lee2010automatic}.

\section{Methodology}
The following is a description of the questionnaire used in the study, the procedure for recruiting participants, and the analysis we used to answer our research questions. Our study protocol was approved by the Institutional Review Board (IRB) of our university. To gather responses regarding our research questions, participants were administered the survey using Qualtrics, and where the language was English. The survey contained quantitative responses to get empirically reasonable insights into privacy and security concerns. It took approximately 26 minutes to complete the survey.

\begin{table*}[h]
\centering
\begin{tabular}{@{}llll@{}}
\toprule
\textbf{No.} & \textbf{Description} & \textbf{Type} & Baseline \\ \midrule
1 & FRT Familiarity: The participants familiarity towards FRT. & Boolean & Familiar \\
2 & Gender: The gender of the participant. & Boolean & Female \\
3 & Age: The age of participants measured in categorical ranges. & Categorical & 25-34 \\
4 & Education: The highest level of educational degree attained by the participant & Categorical & Bachelors degree. \\
5 & Income: The highest level of participants income & Categorical & \$25.000-\$49.999. \\
6 & Occupation: The current job of the participant. & Categorical & Graduate student \\ \bottomrule
\end{tabular}
\caption{The types and descriptions for attributes (covariates) and demographics in the study. Baseline selected based on the majority case. }
\label{table:baseline}
\end{table*}

\subsection{Questionnaire} We conducted the seven-part questionnaire after the consent. In part 1, we have Demographic questions for participants; in part 2, general Awareness and Victim-related questions; in part 3, we have Security Attitudes metrics questions.  We also reused Pew Research Center Questions in part 4, Privacy Concerns, and Security Behavior Intentions scaling questions in part 5 and part 6, correspondingly. Finally, we have asked questions related to Government Trust Questions. 

\textbf{Part 1: Demographic questions} 
Participant demographics were determined by asking standard questions about their age, gender, education, occupation, and income.

\textbf{Part 2: Awareness and Victim related questions} 
Participants were asked questions about their awareness and perception of FRT. These questions were newly created with response options on a 5-point Likert scale. These questions covered the awareness levels of participants to the potential misuse of FRT, and whether they have been a victim of the technology.

\textbf{Part 3: Security Attitudes (SA-6)} 
To measure the security attitudes of Saudis towards FRT, we provide them with previously validated security attitudes (SA6) ~\cite{selfreportedusmeasure}. Here, we have a five-point scale with the “Strongly agree”, “Agree”, “Neutral”, “Disagree”, and “Strongly disagree” along with the “Prefer not to answer” option. Participants who selected “Prefer not to answer” to these questions were not counted during the analysis.

\textbf{Part 4: Pew Research Center Questions} 
In this part, the Pew questions ~\cite{smith2019more} were administered to participants. Participants provided answers about the trust, efficiency, and acceptance of using FRT in different applications. Response options for this part are multiple-choice. Participants who selected “Prefer not to answer” to these questions were not counted during the analysis.

\textbf{Part 5: Privacy Concerns (IUIPC)} 
To measure privacy concerns for Saudis towards FRT, we provide them with the previously validated Internet Users’ Information Privacy Concerns Scale (IUIPC-10) ~\cite{gross2021validity}. Here, we have a five-point scale with “Strongly agree”, “Agree”, “Neutral”, “Disagree”, “Strongly disagree” along with the “Prefer not to answer” option. Participants who selected “Prefer not to answer” to these questions were not counted during the analysis.

\textbf{Part 6: Security Behavior Intentions Scale (SeBIS)} 
To measure the security intentions of Saudis towards FRT, we provide them with previously validated security behavior intention scales of FRT ~\cite{egelman2015scaling}. Here, we have a five-point scale with “Never”, “Rarely”, “Sometimes”, “Often”, and “Always” along with the “Prefer not to answer” option. Participants who selected “Prefer not to answer” to these questions were removed from the sample.

\textbf{Part 7: Government Trust Questions } Participants provide answers for replicate study questions ~\cite{castro2019survey} to evaluate the internet users' opinions on FRT by age and gender, the internet users' opinions on FRT by police, and the internet users' opinions on regulating surveillance cameras and FRT. Response options for this part included a five-point Likert-type scale ranging from strongly agree to strongly disagree. Participants who selected "Neither agree" to these questions were excluded from the sample. Strongly agree and agree options were considered as one group of agreement, and strongly disagree and disagree options were considered as one group of disagreement for our analysis.

\subsection{Recruitment}
To avoid the regulatory challenges of the Saudi Data \& Artificial Intelligence Authority (SDAIA) ~\cite{muasher2021} as per the recommendation of our IRB, we collected data from the local Saudi population in America. The participants were recruited through word-of-mouth referrals and online channels. We have posted our study advertisement into a WhatsApp group comprising Saudi Arabian students who are studying in the US.  Data were collected through a survey instrument targeted at Saudi people above the age of 18 and who identified themselves as Saudis based on nationality. As a token for participation, 1 random participant out of 100 received a 50-dollar Amazon gift card.

\subsection{Analysis}
\hspace{3mm}\textbf{RQ1 analysis:} We present the score for each scale that we asked our participants about, which are Privacy Concerns (IUIPC-8), Security Attitudes (SA-6), and Security Behaviors (SeBIS-30). Then, we used linear regression to analyze the three scales. We set FRT Familiarity, Gender, Age, Income, Education, and Occupations as independent variables for each scale analysis. Every dependent variable was fitted with linear regression using all covariates. Based on majority of the case, we selected the baseline. All Independent variables are listed in Table ~\ref{table:baseline}).

\textbf{RQ2 analysis:} We present the descriptive statistics for each item in Part 4. We compare the response frequency for Likert scale questions for both populations. We used the non-parametric Mann Whitney U test (MWU) since we do not have to assume that the variances and sample size should be equal for both populations.

\textbf{RQ3 analysis:} We present the descriptive statistics for opinions-related questions in Part 7. We compare the response frequency for Likert scale questions for both populations. Like RQ2, we used the non-parametric Mann Whitney U test (MWU) since we do not have to assume that the variances and sample size should be equal for both populations.

\subsection{Data Cleaning}
We received a total of 116 responses. First, we downloaded the survey file from Qualtrics, then we removed entries that did not declare they were from Saudi Arabian and did not provide their name. This is because one of the authors posted an advertisement on the LinkedIn, and some US people filled out the survey without reading the participation requirements. Hence, we removed 22 participants from those categories. As a result, we were left with a total of 94 participants. We then selected surveys that had a 100\% completion rate. Thus, we removed 41 entries of people who completed the survey partially and missed some crucial parts of the survey. Overall, we removed 63 participants in total. 
Then we produced a master file of 53 participants for further analysis. As our survey questions were optional, participants had the freedom not to answer those questions.

\section{Results}
In this section, we first describe our participants' responses to the survey. Then, we present the regression analysis on Security Attitudes, Privacy Concerns, and Security Behaviors. Finally, we analyze Saudis' responses to Pew research center questions and government trust questions toward FRT, and we compare between Saudis and Americans. Results are presented based on an online survey that measures the overall familiarity, trust, and public acceptance, opinions, and efficiency of FRT. Responses to privacy and security questions were expected to show significant differences between gender, but since there was a limited sample size, we only found a couple of significant differences. Also, we noticed that the Saudi sample did not differ significantly from the American sample for the replicated studies.

\subsection{Demographics}
From May 2022 until September 2022, we were able to get 53 valid responses. Of the 53 participants recruited for the study, 17\% of them were between 18–24 years old, 69\% were between 25–34 years old, and 13\% were between 35-44 years. The identified gender distribution for the study was 56.6\% men, and 43.4\% women. More than 74\% of the participants had a Bachelor’s and Master’s degree. 34\% of the participants are graduate students, whereas the rest are distributed between other occupations. Finally, for income, 14\% of Saudis prefer not to reveal their income. Participant characteristics and additional demographic information can be found in Table ~\ref{table:demoresults}.

\subsection{Public Awareness \& Technology Victims}

\textbf{Public Awareness} Our analysis showed that 71.7\% of Saudis have heard about FRT, while 28.3\% have never heard about it. Furthermore, 71.7\% of Saudis agree that they are very aware of the privacy risks concerning FRT, while 11.3\% disagree. Additionally, 68\% of Saudis agree that they are aware that FRT scans can be captured easily and remotely, while 13.2\% of them disagree.

\textbf{Technology Victims} Our participants scored an average of 2.87 (\(\sigma\) = .56, min=1.5, max=4), indicating that they rarely have been a victim of FRT. 73.6\% of Saudis have never been a victim of somebody accessing their facial information to extort them for money, while only 3.8\% of them have been a victim. 73.6\% of Saudis never fell victim to somebody using their facial information under their name, while 5.7\% of them have been a victim. 64.2\% of Saudis have never had their facial information misused for any purpose, such as identity theft, while 5.7\% of them have been a victim. 56.6\% of Saudis have never been a victim of FRT flaws such as false positive identification, while 1.9\% of them faced the risks of error due to flaws in the technology.

\begin{table}[h]
\footnotesize
\centering
\begin{tabular}{@{}cccl@{}}
\toprule
\textbf{} & \textbf{    } & (\textbf{n}) & \multicolumn{1}{r}{\textbf{(\%)}} \\ \midrule
\multirow{2}{*}{\textbf{Gender}} & Male& 30 & 56.6 \\
 & Female & 23 & 43.4 \\ \midrule
\multirow{3}{*}{\textbf{Age}} & 18-24 & 9 & 17.0 \\
 & 25-34 & 37 & 69.8 \\
 & 35-44 & { }{ }7 & 13.2 
 \\ \midrule
\multirow{5}{*}{\textbf{Education}}
 & High School degree & 7 & 13.2 \\
 & Associate degree & 4 & 7.5 \\
 & Bachelor's degree & 21 & 39.6 \\
 & Master's degree & 18 & 34.0 \\
 & Doctoral degree & 2 & 3.8 \\ 
 & Prefer not to answer & 1 & 1.9 \\\midrule

 \multirow{6}{*}{\begin{tabular}[c]{@{}c@{}}{\textbf{Income}} \\ \end{tabular}} & Under \$25,000 & 11 & 20.8 \\
 & \$15,000- \$24,999 & 8 & 15.1 \\
 & \$25,000- \$49,999 & 15 & 28.3 \\
 & \$50,000- \$74,999 & 6 & 11.3 \\
 & \$75,000- \$99,999 & 2 & 3.8 \\
 & \$100,000- \$149,999 & 2 & 3.8 \\
 & \$150,000 or above & 1 & 1.9 \\
 & Prefer not to answer & 8 & 15.1 \\\midrule
\multirow{11}{*}{\textbf{Occupation}} & Art, Writing, or Journalism & 2 & 3.8 \\
 & Business, Management, or Financial & 3 & 5.7 \\
 & Education or Science  & 6 & 11.3 \\
 & Medical & 4 & 7.5 \\
 & IT Professional  & 3 & 5.7 \\
 & Engineer in other fields & 6 & 11.3 \\
 & Service (e.g. retail clerk, server) & 1 & 1.9 \\
 & Skilled Labor (e.g. electrician)e & 1 & 1.9 \\
 & College student & 4 & 7.5 \\
 & Undergraduate student & 4 & 7.5 \\
 & Graduate student & 18 & 34.0 \\
 & Other &  1 & 1.9 \\
 \bottomrule
\end{tabular}
\caption{Participant demographics and their respective frequencies and percentages.}
\label{table:demoresults}
\end{table}

\begin{table}[h]
\centering
\footnotesize
\vspace{-3mm}
\setlength{\tabcolsep}{4pt}
\begin{tabular}{@{}ccclc@{}}
\toprule
\textbf{SA-6 (Baseline)} & \multicolumn{1}{c}{\textbf{\begin{math}\beta\end{math}}} & \multicolumn{1}{c}{\textbf{\begin{math}CI_{95\%}\end{math}}} & \textbf{T-val} & \textbf{\textit{p}-val} \\ \midrule
\multicolumn{1}{l}{{{ \textit{FRT (vs Familiar):}}}} & \multicolumn{1}{l}{} & \multicolumn{1}{l}{} &  & \multicolumn{1}{l}{} \\
Unfamiliar & -0.117 & [-0.522, 0.289]  & \multicolumn{1}{c}{-0.577} & 0.566 \\ \midrule
\multicolumn{1}{l}{{{ \textit{Gender (vs Female):}}}} & \multicolumn{1}{l}{} & \multicolumn{1}{l}{} &  & \multicolumn{1}{l}{} \\
Male & -0.102 &  [-0.470, 0.267]  & \multicolumn{1}{c}{-0.554} & 0.582 \\ \midrule
\multicolumn{1}{l}{{{\textit{Age (vs 25-34):}}}} &  &  &  &  \\
18-24 &  0.020 & [-0.481, 0.521]   & 0.080 & 0.936 \\
35-44 & -0.054 & [-0.610, 0.501]   & -0.195 & 0.846 \\ \midrule
\multicolumn{1}{l}{{{\textit{Education(vs Bachelor's):}}}} & \multicolumn{1}{l}{} & \multicolumn{1}{l}{} &  & \multicolumn{1}{l}{} \\ 
High school degree & 0.270 & [-0.316, 0.855]  & 0.927 & 0.359 \\
Associate degree & 0.157 &  [-0.575, 0.889]  & 0.431 & 0.669 \\
Masters degree &  0.124 &  [-0.307, 0.555] &  0.580 & 0.564\\
Doctoral degree & 0.282 &  [-0.711, 1.275] &  0.571 & 0.571 \\
Prefer not to answer  & -0.968 &  [-2.342, 0.405] &  -1.418 & 0.163 \\ \midrule

\multicolumn{1}{l}{{{\textit{Income(vs\$25,-\$50K ):}}}} &  &  &  &  \\
Under \$25,000  & 0.257 & [-0.251 , 0.764] & 1.018 & 0.314 \\
\$15,000 to \$24,999  & -0.039 & [-0.599, 0.521]  & -0.140 & 0.889 \\
\$50,000 to \$74,999  & 0.100 & [-0.518, 0.718] & 0.326 & 0.746 \\
\$75,000 to \$99,999  & 0.628 & [-0.335, 1.590]  & 1.314 & 0.196 \\
\$100,000 to \$149,999 & 1.211 & [0.249, 2.174] & 2.543 & 0.015* \\
\$150,000 or above & -0.122 & [-1.443, 1.198]  & -0.186 & 0.853 \\ 
Prefer not to answer & -0.206 & [-0        .765, 0.354]  & -0.740 & 0.463 \\\midrule

\multicolumn{1}{l}{{{\textit{Occupation (vs Graduate):}}}} &  &  &  &  \\
Art, Writing, or Journalism & 0.556 & [-0.461, 1.573] &  1.103 & 0.276 \\
Business, Management, or \\Financial & -0.278 &  [-1.129, 0.573]  & -0.659 & 0.513 \\
Education or Science & 0.139 & [-0.504, 0.782]  & 0.436 & 0.665 \\
Medical & 0.222 &  [-0.532, 0.976] & 0.595 & 0.555 \\
Computer Engineering or \\IT Professional & 0.056 & [-1.520, 1.631] & -0.071 & 0.944 \\
Engineer in other fields & 0.611 &  [-0.240, 1.462] & 1.450 & 0.155 \\ 
Service & 0.194 & [-0.449, 0.838] & 0.611 & 0.545 \\
Skilled Labor & -0.111  & [-1.513, 1.291]   & -0.160 & 0.874 \\
College student & 0.556 & [-0.846, 1.957]  & 0.800 &  0.428 \\
Undergraduate student & -0.444 &  [-1.199, 0.310] & -1.190 & 0.241 \\
Other & 0.556 & [-0.846, 1.957] & 0.800 & 0.428 \\ \bottomrule
\end{tabular}
\caption{Final regression table for SA-6. Variables are categorically compared using the majority case as the Baseline.}
\label{tab:demographicsLogistic}
\label{table:securityattitudes}
\end{table}

\begin{table*}[ht]
\tiny
\begin{tabular}{@{}llccccccccc@{}}
\toprule
\textbf{Question} & \textbf{Group} & \textbf{N} & \multicolumn{4}{c}{\textbf{Response(\%)}} & \textbf{p-value} & \textbf{$d_{Cohen}$} & \textbf{Effect Size} & \textbf{Conclusive} \\ \midrule
\textbf{TOPIC(FRT Familiarity)} & \textbf{FACE1 Options} & \multicolumn{1}{l}{} & A lot & A little & Nothing at all & \multicolumn{1}{l}{} &  &  &  &  \\ \midrule
\multirow{2}{*}{\textbf{FACE1}} & Saudi Group & 53 & 20.8 & 67.9 & 11.3 &  & \multirow{2}{*}{0.753} & \multirow{2}{*}{0.05} & \multirow{2}{*}{\begin{tabular}[c]{@{}c@{}}Very small \& \\ not significant\end{tabular}} & \multirow{2}{*}{No} \\
 & U.S. Group & 4260 & 24.3 & 63.0 & 12.6 &  &  &  &  &  \\ \midrule
\textbf{TOPIC(FRT Efficiency)} & \textbf{FACE2 Options} & \multicolumn{1}{l}{} & Very  Effective & Somewhat Effective & Not too much & Not at all &  &  &  &  \\ \midrule
\multirow{2}{*}{\textbf{\begin{tabular}[c]{@{}l@{}}FACE2a\\ (Identify Individuals)\end{tabular}}} & Saudi Group & 53 & 22.6 & 66.0 & 11.3 & 0.0 & \multirow{2}{*}{0.839} & \multirow{2}{*}{0.03} & \multirow{2}{*}{\begin{tabular}[c]{@{}c@{}}Very small \& \\ not significant\end{tabular}} & \multirow{2}{*}{No} \\
 & U.S. Group & 3648 & 24.2 & 61.8 & 12.5 & 1.5 &  &  &  &  \\ \midrule
\multirow{2}{*}{\textbf{\begin{tabular}[c]{@{}l@{}}FACE2b\\ (Identify Gender)\end{tabular}}} & Saudi Group & 48 & 22.9 & 50.0 & 18.8 & 8.3 & \multirow{2}{*}{0.833} & \multirow{2}{*}{\textgreater 0.01} & \multirow{2}{*}{\begin{tabular}[c]{@{}c@{}}Very small \& \\ not significant\end{tabular}} & \multirow{2}{*}{No} \\
 & U.S. Group & 3642 & 18.4 & 55.1 & 21.9 & 4.6 &  &  &  &  \\ \midrule
\multirow{2}{*}{\textbf{\begin{tabular}[c]{@{}l@{}}FACE2c\\ (Identify Race)\end{tabular}}} & Saudi Group & 47 & 21.3 & 34.0 & 34.0 & 10.6 & \multirow{2}{*}{0.104} & \multirow{2}{*}{0.25} & \multirow{2}{*}{\begin{tabular}[c]{@{}c@{}}Small \& \\ not significant\end{tabular}} & \multirow{2}{*}{No} \\
 & U.S. Group & 3623 & 17.7 & 54.7 & 22.9 & 4.8 &  &  &  &  \\ \midrule
\textbf{TOPIC(Trust Scenarios)} & \textbf{FACE3 Options} & \multicolumn{1}{l}{} & A great deal & Somewhat & Not too much & Not at all &  &  &  &  \\ \midrule
\multirow{2}{*}{\textbf{\begin{tabular}[c]{@{}l@{}}FACE3a\\ (Law Enforcement)\end{tabular}}} & Saudi Group & 53 & 11.3 & 47.2 & 28.3 & 13.2 & \multirow{2}{*}{0.103} & \multirow{2}{*}{0.16} & \multirow{2}{*}{\begin{tabular}[c]{@{}c@{}}Very small \& \\ not significant\end{tabular}} & \multirow{2}{*}{No} \\
 & U.S. Group & 3687 & 20.7 & 46.6 & 19.6 & 13.1 &  &  &  &  \\ \midrule
\multirow{2}{*}{\textbf{\begin{tabular}[c]{@{}l@{}}FACE3b\\ (Technology Companies)\end{tabular}}} & Saudi Group & 53 & 18.9 & 28.3 & 38.7 & 15.1 & \multirow{2}{*}{0.046} & \multirow{2}{*}{0.30} & \multirow{2}{*}{\begin{tabular}[c]{@{}c@{}}Small \& somewhat \\ educationally significant\end{tabular}} & \multirow{2}{*}{Somewhat} \\
 & U.S. Group & 3684 & 5.6 & 35.0 & 36.0 & 23.5 &  &  &  &  \\ \midrule
\multirow{2}{*}{\textbf{\begin{tabular}[c]{@{}l@{}}FACE3c\\ (Advertisers)\end{tabular}}} & Saudi Group & 53 & 5.7 & 28.3 & 24.5 & 41.5 & \multirow{2}{*}{0.361} & \multirow{2}{*}{0.17} & \multirow{2}{*}{\begin{tabular}[c]{@{}c@{}}Very small \& \\ not significant\end{tabular}} & \multirow{2}{*}{No} \\
 & U.S. Group & 3688 & 2.5 & 17.4 & 41.1 & 38.9 &  &  &  &  \\ \midrule
\textbf{TOPIC(Acceptance Scenarios)} & \textbf{FACE4 Options} & \multicolumn{1}{l}{} & \multicolumn{1}{l}{Accepted} & \multicolumn{1}{l}{Unaccepted} & \multicolumn{1}{l}{Not sure} &  &  &  &  &  \\ \midrule
\multirow{2}{*}{\textbf{\begin{tabular}[c]{@{}l@{}}FACE4a\\ (Law Enforcement)\end{tabular}}} & Saudi Group & 51 & 49.0 & 31.4 & 19.6 &  & \multirow{2}{*}{0.002} & \multirow{2}{*}{0.32} & \multirow{2}{*}{\begin{tabular}[c]{@{}c@{}}Small \& somewhat \\ educationally significant\end{tabular}} & \multirow{2}{*}{Somewhat} \\
 & U.S. Group & 1876 & 70.7 & 14.8 & 14.5 &  &  &  &  &  \\ \midrule
\multirow{2}{*}{\textbf{\begin{tabular}[c]{@{}l@{}}FACE4b\\ (Apartment Buildings)\end{tabular}}} & Saudi Group & 53 & 41.5 & 37.7 & 20.8 &  & \multirow{2}{*}{0.745} & \multirow{2}{*}{0.07} & \multirow{2}{*}{\begin{tabular}[c]{@{}c@{}}Very small \& \\ not significant\end{tabular}} & \multirow{2}{*}{No} \\
 & U.S. Group & 1830 & 42.2 & 40.3 & 17.5 &  &  &  &  &  \\ \midrule
\multirow{2}{*}{\textbf{\begin{tabular}[c]{@{}l@{}}FACE4c\\ (Employee Attendance)\end{tabular}}} & Saudi Group & 53 & 35.8 & 41.5 & 22.6 &  & \multirow{2}{*}{0.687} & \multirow{2}{*}{0.06} & \multirow{2}{*}{\begin{tabular}[c]{@{}c@{}}Very small \& \\ not significant\end{tabular}} & \multirow{2}{*}{No} \\
 & U.S. Group & 1875 & 34.7 & 48.7 & 16.6 &  &  &  &  &  \\ \midrule
\multirow{2}{*}{\textbf{\begin{tabular}[c]{@{}l@{}}FACE4d\\ (Advertisement)\end{tabular}}} & Saudi Group & 52 & 17.3 & 61.5 & 21.2 &  & \multirow{2}{*}{0.793} & \multirow{2}{*}{0.05} & \multirow{2}{*}{\begin{tabular}[c]{@{}c@{}}Very small \&\\ not significant\end{tabular}} & \multirow{2}{*}{No} \\
 & U.S. Group & 1831 & 16.7 & 64.9 & 18.4 &  &  &  &  &  \\ \bottomrule
\end{tabular}
\caption{Summary of statistical tests results between U.S. and Saudi Groups. }
\label{tab:pew-analysis-table}
\end{table*}
\subsection{Security Attitudes (SA-6)}
We analyzed the responses to the SA-6 security attitude scale questions on the survey. Potential scores on this scale range from 6–30, with higher numbers indicating a more positive attitude toward security behaviors. Overall, all participants scored an average of 21.3 (\(\sigma\) = 3.95, min=11, max=29), meaning that Saudis scored much lower than the average U.S. population sample~\cite{sa6handout,selfreportedusmeasure}. As a result, Saudis are scoring in a lower range compared to Americans for security attitudes on this scale.

\textbf{Regression Model Analysis (SA-6)} 
Our regression model (Table ~\ref{table:securityattitudes}) includes Saudis who are familiar with FRT (Saudi familiar mean= 21.9, Saudi unfamiliar mean = 20.8), but this factor is not significant. Also, we discovered that there is one significant relationship between security attitudes and the independent variable income (covariates).

\textbf{Income covariate:} Saudis whose income falls within the “\$100,000 to \$149,000” range were associated with a 1.21-point increase in positive attitude toward security (\textit{p} = 0.015).

\subsection{Pew Research Center Question Analysis}

\subsubsection{Saudi Public Opinion on Automated FRT} 

In this part of the survey, we replicate a study that has been done on the US population from the Pew Research Center \cite{smith2019more}. We provided our participants with the same questions, and compared the answers between the two populations to analyze the extent of the differences. The broad questions can be found in Table~\ref{table:items} and provide a brief description of the face topics taken from the Pew American Trends Panel\cite{pewfacetopics}. FACE1 references question 1 from the Pew Center Survey and discuss familiarity. FACE2 references questions 2-4 in the Pew Center Survey and discuss efficiency. FACE3 references questions 5-7 in the Pew Center Survey and discuss trust in different scenarios. FACE4 covers questions in the Pew Center Survey 8-11 and cover the acceptance of FRT in different scenarios.

\begin{table}[h]
\footnotesize
\begin{tabularx}{\columnwidth}{l@{\hskip 0.35in}l}
\hline
\multicolumn{1}{l}{\textbf{Item ID}} & \multicolumn{1}{c}{\textbf{Item Text}} \\ \hline
FACE1 & \begin{tabular}[c]{@{}l@{}}How much have you heard or read about the development of\\ automated facial recognition technology that can identify\\ someone based on a picture or video that includes their face?\end{tabular} \\ \hline
FACE2 & \begin{tabular}[c]{@{}l@{}}Based on what you know, how effective do you think facial\\ recognition technology is at the following things?\end{tabular} \\ \hline
FACE3 & \begin{tabular}[c]{@{}l@{}}How much, if at all, do you trust the following groups to use\\ facial recognition technology responsibly?\end{tabular} \\ \hline
FACE4 & \begin{tabular}[c]{@{}l@{}}In your opinion, is it acceptable or unacceptable to use\\ facial recognition technology in the following situations?\end{tabular} \\ \hline
\end{tabularx}
\caption{Items related to facial recognition from the Pew research Center American Trends Panel. The items reference groups of questions from the Pew Research Center Survey.}
\label{table:items}
\end{table}

\subsubsection{Saudi Public Perception on Automated FRT}
In this section, we have provided our participants with the same questions from the Pew research center study that has been done on American people ~\cite{smith2019more}, and we compare the results between Saudis and Americans. We present the descriptive statistics for each section, and use the Mann Whitney U test to determine whether there are any significant difference between the two populations.

\textbf{Public Familiarity}
89\% of Saudis have heard of something related to automated FRT. 21\% said that they have heard a lot about automated FRT, while only 11\% of Saudis have not heard anything at all about facial recognition (Table~\ref{tab:pew-analysis-table}).

\textbf{Public Beliefs on FRT Efficiency}
In general, our participants score an average of 2.31 (\(\sigma\) = .79, min=1, max=4), indicating that they think the use of FRT is not effective when it comes to accurately identifying someone’s race. As shown in Table ~\ref{tab:pew-analysis-table}, Saudis think that the efficiency of FRT at accurately identifying people is 22.6\%. Furthermore, Saudis think that the efficiency of FRT at accurately identifying someone’s race is 18.9\%. In addition, 9.4\% of Saudis think that FRT is not effective at all at accurately identifying someone’s race. We found that there is no significant difference between Saudis' and Americans' opinions on the efficiency of FRT in different scenarios.

\textbf{Public Trust in different scenarios:}
Overall, our participants scored an average of 2.65 (\(\sigma\) = .72, min=1, max=4), indicating that they think the use of FRT by law enforcement agencies, companies, and advertisers is not a great deal. Among the three scenarios, Saudi's trust for the second scenario, “Technology Companies,” is the highest, while it is at its lowest for the “Advertisers” scenario. As depicted in Table ~\ref{tab:pew-analysis-table}, only 5.7\% of Saudis think that it is a great deal to trust advertisers to use FRT in the advertisement scenario, while 41.5\% of them do not trust advertisers at all. A significant difference between Saudis' and Americans trust of advertisers using the FRT has been found (MWU, \textit{p} = 0.01).

\textbf{Public Acceptance in Different Scenarios}
In general, our participants score an average of 1.88 (\(\sigma\) = .48, min=1, max=3), indicating that they are more likely to refuse than accept the use of FRT. 47.2\% of Saudis accept the use of FRT in the scenario of “Law enforcement assessing security threats in public spaces.” While 17\% accept the use of FRT in the scenario “Advertisers seeing how people respond to public ad displays.” From Table ~\ref{tab:pew-analysis-table}, we can see that 60\% of Saudis do not accept the use of FRT in the advertisement scenario, while only 30\% of them do not accept the use of it in the law enforcement scenario. In addition, we found that there is no significant difference between Saudis' and Americans' acceptance of the use of FRT in different scenarios.

\subsection{Privacy Concerns (IUIPC)}
Now, we consider responses to the IUIPC-10, which measures privacy concerns. It is a well-known scale that measures privacy concerns. This scale consists of 10 items divided into three dimensions: three control items (ctrl1, ctrl2, ctrl3), three awareness items (awa1, awa2, awa3), and four collection items (coll1, coll2, coll3, coll4). 

Since IUIPC-8 yielded a statistically significantly better fit to the Vuong test than IUIPC10, we trim ctrl3 and aware3 \cite{gross2021validity}. We converted our 5-point scale to 7-point using a conversion scale mapping~\cite{likertconversion}. Potential scores for IUIPC-8 range from 8-56, with higher scores indicating higher levels of privacy concern. Our participants scored on average of 45.91 (\(\sigma\) = 8.11, min=22, max=56), indicating that they tend to be more privacy sensitive than not.

Our regression model (Table ~\ref{table:privacyconcerns}) includes Saudis who are familiar with FRT (Saudi familiar mean= 47.44, Saudi unfamiliar mean = 45.68); we found that several variables were negatively correlated with positive concerns towards privacy. 

\textbf{Education covariate:} Saudis who have not preferred to reveal their education degree are associated with a 2.89-point decrease in positive privacy concerns (\textit{p} = .002).

\begin{table}[H]
\centering
\resizebox{.79\textwidth}{!}{%
\setlength{\tabcolsep}{4pt}
\begin{tabular}{@{}ccclc@{}}
\toprule
\textbf{IUIPC-8 (Baseline)} & \multicolumn{1}{c}{\textbf{\begin{math}\beta\end{math}}} & \multicolumn{1}{c}{\textbf{\begin{math}CI_{95\%}\end{math}}} & \textbf{T-val} & \textbf{\textit{p}-val} \\ \midrule
\multicolumn{1}{l}{{{ \textit{FRT (vs Familiar):}}}} & \multicolumn{1}{l}{} & \multicolumn{1}{l}{} &  & \multicolumn{1}{l}{} \\
Unfamiliar & -0.226 & [-0.856, 0.404]  & \multicolumn{1}{c}{-0.719} & 0.475 \\ \midrule
\multicolumn{1}{l}{{{ \textit{Gender (vs Female):}}}} & \multicolumn{1}{l}{} & \multicolumn{1}{l}{} &  & \multicolumn{1}{l}{} \\
Male & 0.325 &  [-0.243, 0.894]  & \multicolumn{1}{c}{1.148} & 0.256 \\ \midrule
\multicolumn{1}{l}{{{\textit{Age (vs 25-34):}}}} &  &  &  &  \\
18-24 &  -0.322 & [-1.095, 0.451]   & -0.837 & 0.407 \\
35-44 & 0.148 & [-0.709, 1.005]   &  0.347 & 0.730 \\ \midrule
\multicolumn{1}{l}{{{\textit{Education(vs Bachelor's):}}}} & \multicolumn{1}{l}{} & \multicolumn{1}{l}{} &  & \multicolumn{1}{l}{} \\

High school degree & 0.589 & [-0.167, 1.346]  & 1.567 & 0.124 \\
Associate degree & -0.893 &  [-1.838, 0.053]  & -1.900 & 0.064 \\
Masters degree &  0.524 &  [-0.033, 1.080] &  1.893 & 0.065\\
Doctoral degree & -0.830 &  [-2.113, 0.452] &  -1.303 & 0.199 \\
Prefer not to answer  & -2.893 &  [-4.667, -1.119]  &  -3.281 & 0.002* \\ \midrule

\multicolumn{1}{l}{{{\textit{Income(vs\$25,-\$50K ):}}}} &  &  &  &  \\
Under \$25,000  & -0.848 & [-1.606, -0.089]  & -2.251 & 0.029* \\
\$15,000 to \$24,999  & -0.419 & [-1.255, 0.418]  & -1.008 & 0.319 \\
\$50,000 to \$74,999  & -0.387 & [-1.311, 0.536]  & -0.845 & 0.402 \\
\$75,000 to \$99,999  & -0.825 & [-2.264, 0.614]  & -1.155 & 0.254 \\
\$100,000 to \$149,999 & -0.325 & [-1.764, 1.114] & -0.455 & 0.651 \\
\$150,000 or above & -2.700 & [-4.674, -0.726]  & -2.755 & 0.008* \\ 
Prefer not to answer & 0.316 & [-0.521, 1.152]  & 0.760 & 0.451 \\\midrule

\multicolumn{1}{l}{{{\textit{Occupation (vs Graduate):}}}} &  &  &  &  \\
Art, Writing, or Journalism & -1.375 &  [-2.857, 0.107  ] & -1.873 & 0.068 \\
Business, Management, or \\Financial & -0.667 & [-1.907, 0.574 ]	& -1.086 &	0.284 \\
Education or Science & -0.604 &	[-1.542, 0.333] & -1.301	& 0.200	 \\
Medical & -0.156 &	[-1.256, 0.943] & -0.287	& 0.776 \\
Computer Engineering or \\IT Professional & -0.250 & [-1.490, 0.990] & -0.407	& 0.686	 \\
Engineer in other fields & -0.875 &	[-1.813, 0.063] & -1.885	& 0.067	 \\ 
Service & -1.625 &	[-3.668, 0.418] & -1.606 &	0.116 \\
Skilled Labor & -2.375 & [-4.418, -0.332] & -2.347 & 0.024* \\
College student & -1.250 & [-2.349, -0.151] & -2.296	& 0.027*	 \\
Undergraduate student & -0.469 &	[-1.568, 0.631] & -0.861	& 0.394 \\
Other & -0.875 &	[-2.918, 1.168]	& -0.865	 & 0.392 \\ \bottomrule

\end{tabular}
}
\caption{Final regression for IUIPC-8. Variables are categorically compared using the majority case as the Baseline.}
\label{table:privacyconcerns}
\end{table}

\begin{table*}[h]
\resizebox{\textwidth}{!}{%
\begin{tabularx}{1.25\textwidth}{@{}llllll@{}}
\toprule
\multicolumn{1}{c}{\textbf{Proposition}} & \multicolumn{1}{c}{\textbf{\begin{tabular}[c]{@{}c@{}}Agree/\\ Disagree\end{tabular}}} & \multicolumn{1}{c}{\textbf{\begin{tabular}[c]{@{}c@{}}18-34\\ SA/US\end{tabular}}} & \multicolumn{1}{c}{\textbf{\begin{tabular}[c]{@{}c@{}}34-54\\ SA/US\end{tabular}}} & \multicolumn{1}{c}{\textbf{\begin{tabular}[c]{@{}c@{}}Male\\ SA/US\end{tabular}}} & \multicolumn{1}{c}{\textbf{\begin{tabular}[c]{@{}c@{}}Female\\ SA/US\end{tabular}}} \\ \midrule
Agree or disagree? The government should strictly limit the use of FRT. & \begin{tabular}[c]{@{}l@{}}Agree\\ Disagree\end{tabular} & \begin{tabular}[c]{@{}l@{}}43.2\%-29.8\%\\ 36.4\%-38.9\%\end{tabular} & \begin{tabular}[c]{@{}l@{}}50.0\%-25.7\%\\ 33.3\%-44.6\%\end{tabular} & \begin{tabular}[c]{@{}l@{}}35.7\%-29.4\%\\ 39.3\%-44.1\%\end{tabular} & \begin{tabular}[c]{@{}l@{}}54.5\%-23.0\%\\ 31.8\%-45.6\%\end{tabular} \\ \midrule
\begin{tabular}[c]{@{}l@{}}Agree or disagree? The government should strictly limit the use of FRT even if\\ it means stores can’t use it to reduce shoplifting.\end{tabular} & \begin{tabular}[c]{@{}l@{}}Agree\\ Disagree\end{tabular} & \begin{tabular}[c]{@{}l@{}}40.9\%-27.2\%\\ 40.9\%-43.4\%\end{tabular} & \begin{tabular}[c]{@{}l@{}}57.1\%-22.7\%\\ 14.3\%-48.5\%\end{tabular} & \begin{tabular}[c]{@{}l@{}}37.9\%-26.7\%\\ 41.4\%-48.0\%\end{tabular} & \begin{tabular}[c]{@{}l@{}}50.0\%-21.1\%\\ 31.8\%-50.1\%\end{tabular} \\ \midrule
\begin{tabular}[c]{@{}l@{}}Agree or disagree? The government should strictly limit the use of FRT even if\\ it means airports can’t use it to speed up security lines.\end{tabular} & \begin{tabular}[c]{@{}l@{}}Agree\\ Disagree\end{tabular} & \begin{tabular}[c]{@{}l@{}}38.6\%-24.1\%\\ 47.7\%-49.2\%\end{tabular} & \begin{tabular}[c]{@{}l@{}}28.6\%-18.5\%\\ 42.9\%-53.2\%\end{tabular} & \begin{tabular}[c]{@{}l@{}}37.9\%-23.1\%\\ 51.7\%-53.8\%\end{tabular} & \begin{tabular}[c]{@{}l@{}}36.4\%-16.9\%\\ 40.9\%-54.9\%\end{tabular} \\ \midrule
\begin{tabular}[c]{@{}l@{}}Agree or disagree? The government should strictly limit the use of FRT even if\\ it comes at the expense of public safety.\end{tabular} & \begin{tabular}[c]{@{}l@{}}Agree\\ Disagree\end{tabular} & \begin{tabular}[c]{@{}l@{}}29.5\%-20.2\%\\ 54.5\%-52.0\%\end{tabular} & \begin{tabular}[c]{@{}l@{}}28.6\%-18.0\%\\ 57.1\%-52.5\%\end{tabular} & \begin{tabular}[c]{@{}l@{}}20.7\%-22.7\%\\ 62.1\%-53.3\%\end{tabular} & \begin{tabular}[c]{@{}l@{}}40.9\%-14.1\%\\ 45.5\%-56.3\%\end{tabular} \\ \bottomrule
 \end{tabularx}
 }
\caption{Saudis’ and Americans’(US) opinions on (FRT), by age and gender.}
\label{tab:ageandgender}
\end{table*}

\textbf{Income covariate:} Saudis whose income falls “under \$25,000” were associated with a 0.8-point decrease in positive concerns towards privacy (\textit{p} = 0.029), and Saudis whose income is in “\$150,000 or above” range were associated with a 2.7-point decrease in positive concerns towards privacy (\textit{p} = 0.008).

\textbf{Occupation covariate:} Saudis who have an occupation in skilled labor are associated with a 2.38-point decrease in positive privacy concerns (\textit{p} = 0.024), and Saudis who are college students are associated with a 1.25-point decrease in positive privacy concerns (\textit{p} = .027).

\subsection{Security Behavior (SeBIS-30)}
First, we analyzed the responses to the Security Behavior Intention Scale (SeBIS-30) questions on the survey, where higher numbers indicate a more positive attitude toward security behaviors. The SeBIS-30 questionnaire examines whether cultural differences influence end-users security behavior. Overall, all participants scored an average of 95.67 (\(\sigma\) = 14.66, min=72, max=159).

\textbf{Regression Model Analysis (SeBIS-30)} 
Our regression model (Table~\ref{table:securitybehaviors}) includes Saudis who are familiar with FRT (Saudi familiar mean= 96, Saudi unfamiliar mean = 95.4), but this factor is also not significant. Also, we found that there are two significant differences between security behaviors and the independent variables income and occupation (covariates).

\textbf{Income covariate:} Saudis whose income is in the “\$150,000 or above” range was associated with a 2.15-point increase in positive behaviors toward security (\textit{p} $<$ 0.001).

\textbf{Occupation covariate:} Saudis who are college students were associated with a 1.25-point decrease in positive behaviors toward security (\textit{p} = 0.027).

\begin{table}[h]
\centering
\footnotesize
\setlength{\tabcolsep}{4pt}
\begin{tabular}{@{}ccclc@{}}
\toprule
\textbf{SeBIS (Baseline)} & \multicolumn{1}{c}{\textbf{\begin{math}\beta\end{math}}} & \multicolumn{1}{c}{\textbf{\begin{math}CI_{95\%}\end{math}}} & \textbf{T-val} & \textbf{\textit{p}-val} \\ \midrule
\multicolumn{1}{l}{{{ \textit{FRT (vs Familiar):}}}} & \multicolumn{1}{l}{} & \multicolumn{1}{l}{} &  & \multicolumn{1}{l}{} \\
Unfamiliar & 0.037 &  [-0.264, 0.339] & \multicolumn{1}{c}{0.249} & 0.804 \\ \midrule
\multicolumn{1}{l}{{{ \textit{Gender (vs Female):}}}} & \multicolumn{1}{l}{} & \multicolumn{1}{l}{} &  & \multicolumn{1}{l}{} \\
Male & -0.034 &   [-0.308, 0.240]   & \multicolumn{1}{c}{-0.247} & 0.806 \\ \midrule
\multicolumn{1}{l}{{{\textit{Age (vs 25-34):}}}} &  &  &  &  \\
18-24 &   0.281   &   [-0.080, 0.641] & 1.565 &	0.124	 \\
35-44 & 0.222    &   [-0.178, 0.622] &	     1.113 &	0.271 \\ \midrule
\multicolumn{1}{l}{{{\textit{Education(vs Bachelor's):}}}} & \multicolumn{1}{l}{} & \multicolumn{1}{l}{} &  & \multicolumn{1}{l}{} \\ 

High school degree &  -0.111	&    [-0.553, 0.330]   &    -0.506 & 0.615	 \\
Associate degree & 	  0.253	&    [-0.299, 0.805]    &     0.923 & 0.361	 \\
Masters degree &  0.027	&    [-0.298, 0.352]   &     0.169	& 0.867\\
Doctoral degree & 0.087	 &   [-0.662, 0.835]   &     0.233 & 0.817	 \\
Prefer not to answer  &  0.437	 &   [-0.599, 1.472]   &   0.848  &  0.401 \\ \midrule

\multicolumn{1}{l}{{{\textit{Income(vs\$25,-\$50K)}:}}} &  &  &  &  \\

Under \$25,000 &  -0.016	&     [-0.343, 0.311] &	-0.101 &	 0.920\\

\$15,000 to \$24,999  & -0.005    &    [-0.366, 0.356] &	-0.028 & 0.978 \\

\$50,000 to \$74,999  & -0.152    &    [-0.550, 0.246] &	-0.771 & 0.445 \\

\$75,000 to \$99,999  &   0.020	 &    [-0.600, 0.640] &	0.065 &	0.948 \\
\$100,000 to \$149,999 & 0.270	  &   [-0.350, 0.890] &	0.877	& 0.385 \\
\$150,000 or above & 2.153	  &   [1.303, 3.004] &	5.098 & 0.000*	 \\ 
Prefer not to answer &  0.078  &     [-0.282, 0.439] &	0.438	& 0.664 \\\midrule
\multicolumn{1}{l}{{\textit{\underline {Occupation (vs Graduate)}:}}} &  &  &  &  \\
Art, Writing, or Journalism & 0.583 &	[-0.143, 1.309]	&  1.623	& 0.112 	 \\

Business, Management  & -0.200 &	[-0.807, 0.407]	& -0.665	 & 0.510 \\

Education or Science  & -0.283 &	[-0.742, 0.176]	& -1.246	& 0.220 \\

Medical & 0.150 &	[-0.388, 0.688]	& 0.563	& 0.577	 \\

Computer Engineering & 0.122 &	[-0.485, 0.730]    & 0.406	& 0.687		 \\

Engineer in other fields& -0.100 &	[-0.559, 0.359]	 & -0.440	& 0.662 \\ 

Service & -0.467 &	[-1.467, 0.534]	& -0.942	& 0.352	 \\

Skilled Labor & 0.200 & [-0.800, 1.200]	& 0.404	& 0.689	 \\

College student & -1.250 & [-2.349, -0.151] & -2.296	& 0.027*	 \\

Undergraduate student & -0.333 & [-0.872, 0.205]	& -1.251	& 0.218	 \\

Other & -0.800 & [-1.800, 0.200]	& -1.615	& 0.114 \\ \bottomrule

\end{tabular}
\caption{Final regression table for SeBIS. Variables are categorically compared using the majority case as the Baseline.}
\label{table:securitybehaviors}
\end{table}

\begin{table*}[hbt!]
\resizebox{.99\textwidth}{!}{%
\begin{tabular}{lcccccccc}
\hline
\multicolumn{1}{c}{\textbf{Proposition}} & \textbf{\begin{tabular}[c]{@{}c@{}}Saudi\\ Agree\end{tabular}} & \textbf{\begin{tabular}[c]{@{}c@{}}US\\ Agree\end{tabular}} & \textbf{\begin{tabular}[c]{@{}c@{}}Saudi\\ Disagree\end{tabular}} & \textbf{\begin{tabular}[c]{@{}c@{}}US\\ Disagree\end{tabular}} & \textbf{p-value} & \textbf{Cohen’s d} & \textbf{Effect Size} & \textbf{Conclusive} \\ \hline
\textbf{\underline{Topic: Opinions on facial recognition technology}} &  &  &  &  &  &  &  &  \\
\begin{tabular}[c]{@{}l@{}}Agree or disagree? - The government should strictly limit the use of facial \\ recognition technology.\end{tabular} & 44.0\% & 26.2\% & 36.0\% & 44.9\% & 0.019 & 0.084 & \begin{tabular}[c]{@{}c@{}}Very Small \& somewhat\\ educationally significant\end{tabular} & Somewhat \\
\begin{tabular}[c]{@{}l@{}}Agree or disagree?- The government should strictly limit the use of facial \\ recognition technology even if it means stores can’t use it to reduce shoplifting.\end{tabular} & 43.1\% & 23.8\% & 37.3\% & 49.1\% & 0.005 & 0.096 & \begin{tabular}[c]{@{}c@{}}Very Small \& somewhat\\ educationally significant\end{tabular} & Somewhat \\
\begin{tabular}[c]{@{}l@{}}Agree or disagree? - The government should strictly limit the use of facial \\ recognition technology even if it means airports can’t use it to speed up security \\ lines.\end{tabular} & 37.3\% & 20.0\% & 47.1\% & 54.3\% & 0.012 & 0.08 & \begin{tabular}[c]{@{}c@{}}Very Small \& somewhat\\ educationally significant\end{tabular} & Somewhat \\
\begin{tabular}[c]{@{}l@{}}Agree or disagree? - The government should strictly limit the use of facial \\ recognition technology even if it comes at the expense of public safety.\end{tabular} & 29.4\% & 18.3\% & 54.9\% & 54.8\% & 0.141 & 0.046 & \begin{tabular}[c]{@{}c@{}}Very small \&\\ not significant\end{tabular} & No \\
\textbf{\underline{Topic: Opinions on use of FRT by police}} &  &  &  &  &  &  &  &  \\
\begin{tabular}[c]{@{}l@{}}Agree or disagree? Police departments should be allowed to use facial recognition\\ technology to help find suspects if the software is correct 80\% of the time.\end{tabular} & 54.9\% & 39.3\% & 25.5\% & 32.1\% & 0.091 & 0.061 & \begin{tabular}[c]{@{}c@{}}Very small \&\\ not significant\end{tabular} & No \\
\begin{tabular}[c]{@{}l@{}}Agree or disagree? Police departments should be allowed to use facial recognition \\ technology to help find suspects if the software is correct 90\% of the time.\end{tabular} & 62.0\% & 47.3\% & 22.0\% & 25.0\% & 0.256 & 0.034 & \begin{tabular}[c]{@{}c@{}}Very small \&\\ not significant\end{tabular} & No \\
\begin{tabular}[c]{@{}l@{}}Agree or disagree? Police departments should be allowed to use facial recognition \\ technology to help find suspects if the software is correct 100\% of the time.\end{tabular} & 74.0\% & 59.4\% & 12.0\% & 16.1\% & 0.242 & 0.034 & \begin{tabular}[c]{@{}c@{}}Very small \&\\ not significant\end{tabular} & No \\
\textbf{\underline{Topic: Opinions on Surveillance cameras}} &  &  &  &  &  &  &  &  \\
\begin{tabular}[c]{@{}l@{}}Agree or disagree? The government should strictly limit the use of surveillance \\ cameras.\end{tabular} & 44.2\% & 36.2\% & 23.1\% & 29.4\% & .215 & 0.047 & \begin{tabular}[c]{@{}c@{}}Very small \&\\ not significant\end{tabular} & No \\
\begin{tabular}[c]{@{}l@{}}Agree or disagree? The government should strictly limit the use of surveillance \\ cameras even if it means stores can’t use them to reduce shoplifting.\end{tabular} & 32.7\% & 18.2\% & 44.2\% & 58.8\% & 0.006 & 0.092 & \begin{tabular}[c]{@{}c@{}}Very Small \& somewhat\\ educationally significant\end{tabular} & Somewhat \\
\begin{tabular}[c]{@{}l@{}}Agree or disagree? - The government should strictly limit the use of surveillance\\ cameras even if it comes at the expense of public safety.\end{tabular} & 23.5\% & 17.9\% & 56.9\% & 58.6\% & .426 & 0.024 & \begin{tabular}[c]{@{}c@{}}Very small \&\\ not significant\end{tabular} & No \\ \hline
\end{tabular}
}
\caption{Government Trust Questions Analysis: Center for Data Innovation results compared to Saudi Population.}
\label{tab:cdi-analysis-table}
\end{table*}

\subsection{Government Trust Questions Analysis}
\textbf{Saudis’ and Americans’ Opinions on FRT}
We found that 44.0\% of Saudis think that the government should strictly limit the use of FRT, while 26.2\% of Americans support limiting the use of FRT. Likewise, only 43.1\% of Saudis want the government to limit the use of FRT even if it would prevent stores from using this technology to stop shoplifting, while 23.8\% of Americans would agree to such a tradeoff. As highlighted in Table~\ref{tab:cdi-analysis-table}, 37.3\% of Saudis want the government to limit the use of FRT even if it means that the airport can’t use FRT to speed up security lines, while 20\% of Americans would agree to such a tradeoff (Table ~\ref{tab:cdi-analysis-table}). Across the four propositions, we found that the Saudi sample had a higher average agreement that the government should strictly limit the use of FRT. In conclusion, there is a difference in perspectives between the two populations regarding the limiting of FRT technology.

\textbf{Saudis’ and Americans’ Opinions on FRT by Age and Gender}
Overall, we concluded that there is no significant difference between male and female Saudis who agree or disagree with the strict limitation of FRT use by the government. For Saudis, disagreement is prevalent between the 18-34 age group on the limitation of using FRT in different scenarios. As a result, both populations share a similar opinion on this proposition. 35.7\% of men are less likely to agree on the limited use of FRT, especially at the expense of public safety, compared to 54.5\% of women. Also, 37.9\% of men are more likely to agree on the limited use of FRT, especially when it comes to reducing shoplifting and speeding up security lines, compared to 50.0\% of women. For Americans, there were differences in these opinions based on age, with older Americans being less likely to disagree with government limitations on the use of FRT. For example, 52\% of 18- to 34-year-olds disagree with limitations that come at the expense of public safety, compared to 54.5\% of Saudis respondents. We noticed that the Saudi women group were much more likely to support limiting the use of FRT than US women in all the scenarios in (Table ~\ref{tab:ageandgender}).

\textbf{Saudis’ and Americans’ Opinions on the Use of FRT by Police Departments}
After asking Saudi participants whether police departments should be allowed to use FRT to help find suspects, the number of Saudis who agree and support using this technology increased depending on its accuracy. For Saudis, if the software is accurate 80\% of the time, 54.9\% of Saudis agree with using it, whereas 25\% disagree. In contrast, if the software is accurate 90\% of the time, 62.0\% of the participant respondents agree with using it, and 22.0\% disagree. However, if the software is accurate 100\% of the time, 74.0\% of the participants agree with using it, while 12.0\% of them disagree. When U.S. participants were asked whether police departments should be allowed to use FRT to help find suspects, they found that Americans who agree and support using this technology also increased depending on its accuracy. For Americans, if the software is accurate 80\% of the time, 39\% of Americans agree with using it, whereas 32\% disagree. If the software is accurate 90\% of the time, 47\% of the participants respondents agree with using it, and 25\% disagree. However, if the software is accurate 100\% of the time, 59\% of the participants agree with using it, and 16\% disagree (Table~\ref{tab:cdi-analysis-table}). Despite an FRT accuracy of 100\%, approximately 12\% of the two populations do not believe that FRT should be used by the police to find suspects.

\textbf{Saudis’ Opinions on Regulating Surveillance Cameras and FRT}
When participants were asked whether the government should limit the use of surveillance cameras, Saudis were more likely to support such limitation. The support for limiting surveillance cameras when it comes to public safety, 23.5\% of Saudis agree to limit the use of surveillance cameras, and 32.7\% would agree with using facial recognition. Americans were more likely to support limiting the use of surveillance cameras 36.2\% than FRT 26.2\%. The support for limits on surveillance cameras, even if it is going to reduce shoplifting, drops from 44.2\% to just 23.5\% by the Saudi group, and the support for limiting the use of facial recognition drops slightly from 44.0\% to 43.1\%. Alternatively, when it comes to public safety, then 17.9\% of Americans agree to limit the use of surveillance cameras, and 18.3\% would agree to use FRT. These findings have been outlined in (Table~\ref{tab:cdi-analysis-table}).

\subsubsection{Findings from the comparison between the two populations}

We compare our participants to the participants from the Center of Data Innovation study \cite{castro2019survey}. Since we do not have to assume that the variances and sample size should be equal for the two populations, we chose to apply the Mann Whitney U Test to find if there are any significant differences between Saudis and Americans.

We found a couple of significant differences between the two populations for the opinions on limiting the use of FRT(MWU, \textit{p} = 0.019). This indicates that Saudis are more in agreement on the limitation of the use of FRT than Americans in some scenarios. In other words, Saudis tend to be more in agreement than Americans when it comes to limiting the use of FRT, even when used to reduce shoplifting (MWU, \textit{p} = 0.005) or if it means that airports can't use it to speed up security lines(MWU, \textit{p} = 0.012). Also, for their opinions on surveillance cameras, Saudis tend to be less in agreement than Americans when it comes to limiting the use of surveillance cameras, even when used to reduce shoplifting (MWU, \textit{p} = 0.006). For the remaining questions, they shared a similar opinion on the use of FRT by the government.

\textbf{Sample Size and Statistical Power}
After our initial experiment, we performed a post hoc power analysis to determine whether our non-significant results were due to the modest sample size (N=53) of our Saudi population. With power (1-$\beta$) set at 0.95 and $\alpha$ = 0.05 using a two-tailed means statistical comparison, we found a between group effect size of (0.5) for the current sample size used in this study. Thus, with our current population we are able to detect large and medium effect sizes with 95\% power when applying the Mann Whitney U Test. 

\section{Discussion}
In our research, we aimed to understand the perceptions and attitudes of people with respect to FRT. The results of our study have some implications, and we acknowledge that a larger data set may be necessary to draw conclusions on the broader Saudi population.

Due to skewness of the data collected, the results might be biased toward Saudis who are younger, and more educated. Accordingly, the conclusions we draw are not representative of all Saudis. Therefore, our findings cannot be assumed to represent the public perceptions in the Kingdom of Saudi Arabia.

To answer our RQ1, we found that there were some differences in security and privacy concerns and behaviors. One of our findings states that Saudis score much lower than Americans when it comes to security attitudes. While we do not claim that this finding will be the same for a representative sample of Saudis, it does illustrate a potential nuance in the cultural differences between the two groups. Based on the results, we found a similar change in the opinions of FRT usage by police based on its efficacy.

In regard to RQ2, another finding showed that Saudis are more likely than Americans to agree that the government should strictly limit the use of FRT. As mentioned earlier, this finding cannot be claimed as that of the public in Saudi Arabia. Both populations share the same order of acceptance regarding trust in this technology in four different scenarios, whereby they trust the law enforcement scenario the most and the advertiser’s scenario the least. This could be intuitively similar to a representative sample of Saudis and other populations.

In our RQ3, we asked to what extent does Gender and Age impact security behavior and privacy concerns. But we did not find any impact on security behaviors, neither privacy concerns. We do not guarantee this result will be the same for Female Gender who are living in the Kingdom. In several studies, gender has shown a difference. In addition, most women who are living in the Kingdom of Saudi Arabia are wearing Hijab and large number of them are covering their faces. Women who are residing in the United States might not be covering their faces and their responses might not be the same when they go back to the Kingdom regarding privacy concerns.

 \textbf{Recommendations} 
In regard to the acceptance rates of FRT, we found that opinions improved along with the accuracy of the technology. This bolsters the case made by prior research that false identification of FRT by law enforcement is a significant concern ~\cite{zhang2021facial}. As a result, more research could be done to improve the perception of FRT as a reliable technology. 

Regarding the regulation of surveillance cameras, we found that the Saudi population was more likely to support limitations on the use of FRT. Although this is not a topic explored in this study, future work could explore follow-up questions to explore this difference. Additionally, our results show the public beliefs in the efficiency of FRT is so low across multiple use scenarios. This is an issue that should be addressed by both public and private institutions detailing how they use FRT. More should be done to make information detailing the accuracy of FRT in different scenarios available.

\textbf{Future Work} While our work includes information on the differences between Saudi and American opinions on FRT in several contexts, there are many left to explore. FRT is expanding into many fields such as hospitality services and financial sectors. Understanding the differences in the perceptions of FRT in different cultural contexts could provide important insights into how these concerns should be addressed in the future.

\textbf{Limitations} Our study includes certain limitations that are common for this type of research. First, since we collected only 53 responses and due to the regulatory challenges of the Saudi Data \& Artificial Intelligence Authority (SDAIA), we obtained results from a modest sample that does not statistically represent the entire Saudi population. Instead, our study focused on the Saudi population residing in America. Additionally, while our survey respondents were residing within the US, we did not consider how long they have been currently residing in the United States as a factor in our analysis. Differences in the length of their stay in the United States may affect their responses to these topics.  However, our sample still provides valuable insights into the perceptions of FRT. 

Second, most of our participants who completed the survey were students and might be more educated when compared to the average person in Saudi Arabia, and we compared the results for the Pew scales to a more representative sample of the U.S. population. Thus, our study outcomes cannot be generalized to Saudis in Saudi Arabia.

Third, survey responses were only collected from Saudis who currently reside in the United States to avoid confounds related to availability and popularity, as well as cultural differences. Nevertheless, a more representative sample of Saudis is recommended and could be more valuable for further research.

Lastly, we noticed an inconsistency with the responses between a Pew Research Center question and a general question regarding FRT in our survey. This may be due to the Pew Research Center question being double barreled. In particular, we found 71.7\% of participants who reported not hearing about FRT in the General Questions, while expressing that 89.9\% had heard or read about the development of automated FRT in the Pew Research Center Question. The Research Center question being asked later in the survey, the different in complexity of the question, or the extra information about FRT embedded in the question itself giving additional context to FRT technology may have caused participants to indicate they had more understanding of the technology than in the previous response. 
\section{Conclusion}
FRT is a growing area of mass surveillance that raises privacy issues and is seen as an increasing menace in today's society. Despite there are several studies on understanding the security and privacy concerns of FRT in the context of western cultures, there have been no prior studies for the Muslim-majority countries such as Saudi Arabia, which has unique cultures and norms.  To gain insights into the security concerns and attitudes toward the FRT of Saudi Arabian, we designed and conducted an online survey with 53 participants.  Our study sheds light on whether Saudis and Americans have meaningful differences in behavioral attitudes and concerns regarding the privacy and security of FRT that they could realistically encounter as part of their everyday activities. We used previously validated metrics to answer our research questions and found a couple of differences between Saudis and Americans. In terms of security attitudes and privacy concerns, Saudis are more likely to score lower than Americans. When we compared our sample to a U.S. representative sample, we found that the acceptance of FRT by Saudis and Americans in various scenarios does not vary significantly. In addition, more Saudis than Americans believe FRT should be strictly limited by the government. Moreover, future studies on FRT can be guided by our findings, and efforts should be made to ensure the privacy of individuals as FRT advances. Our study findings have implications for policymakers, researchers, and the public regarding the use of FRT in Saudi Arabia. 
\section{Acknowledgement}

The authors are grateful to all the participants who participated in our research. We appreciate PST'23 anonymous reviewers for their thoughtful feedbacks and comments and helped us to improve final version of the paper. We also want to thank Soha Khoso for proofreading the draft version of the paper.

\bibliographystyle{plain}
\bibliography{main}
\end{document}